\documentclass{elsart}

\usepackage{natbib}
\usepackage{psfig}
\usepackage{amssymb}

\begin{document}
\begin{frontmatter}

\title{{\bf A puzzling event during the X-ray emission of the binary system GX 1+4}}

\author[OAR,RM2]{N. Rea\thanksref{now}},
\ead{n.rea@sron.nl}
\author[OAR]{L. Stella},
\author[OAR]{G.L. Israel} ,
\author[RM3]{G. Matt}, 
\author[MSS]{S. Zane}, 
\author[PAL]{A. Segreto} and
\author[EST]{T. Oosterbroek}
     \thanks[now]{Current address: SRON--National Institute for Space Research, Sorbonnelaan 2, 3584 CA, Utrecht, The Netherlands}

\address[OAR]{INAF - Osservatorio Astronomico di Roma, Italy}
\address[RM2]{Physics Department, Universit\`a degli Studi di Roma ``Tor Vergata'', Italy}
\address[RM3]{Physics Department, Universit\`a degli Studi di Roma Tre, Italy}
\address[MSS]{MSSL, University College of London, UK}
\address[PAL]{IASF-- Consorzio Nazionale delle Ricerche, Palermo, Italy}
\address[EST]{Research and Scientific Support Department of ESA-ESTEC, The Netherlands}

\begin{abstract}
  We report on a long X-ray observation of the slow-rotating
  binary pulsar GX\,1+4. {\it BeppoSAX}\, observed, in the
  0.1--200\,keV energy range, an event in which the source flux
  dropped for almost a day, and then recovered. During this event only
  the high-energy emission was found to be pulsed and the pulsations
  were shifted in phase of $\sim 0.2$. The spectrum during the event
  was well fitted by a Compton-reflection model. A broad iron line at
  $\sim6.55$\,keV was present outside of the event, where instead two
  narrow emission lines at $\sim6.47$\,keV and $\sim7.05$\,keV were
  detected. The pulse profile was highly variable as a function of
  both energy and time. We interpret this low-flux event as an
  occultation of the direct X-ray emission, due to the increase of a
  torus-like accretion disk; we then discuss similarities between this
  source and the recently discovered highly absorbed {\it INTEGRAL}
  sources.
\end{abstract}
 
\begin{keyword}
Neutron star \sep pulsar  \sep binary system  \sep GX\,1+4 \sep Compton reflection
\end{keyword}
\end{frontmatter}

\section{Introduction}
%%%%%%%%%%%%%%%%%%%%%%%%%%%%%%%%%%%%%%%%%%
\begin{figure}[t]
\centerline{\psfig{figure=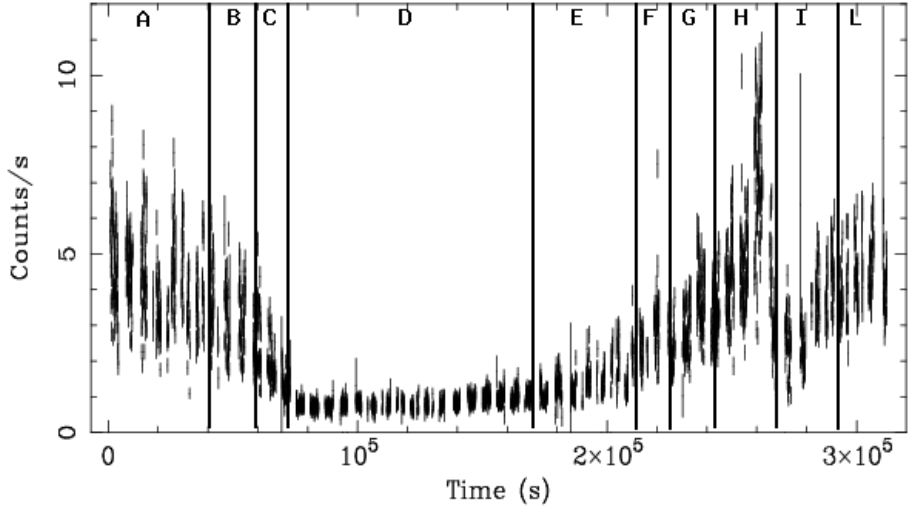,width=12cm}}
\caption{lightcurve (1.65--10\,keV) of the {\it BeppoSAX} 2000 
observation  of GX\,1+4 with superimposed the time intervals used in the analysis.}
\end{figure} 
%%%%%%%%%%%%%%%%%%%%%%%%%%%%%%%%%%%%%%%

GX\,1+4\,\, is an X-ray binary system harboring a $\sim$130\,s pulsar
([13, 10, 8]) accreting mass from a red giant companion of class M5 III
(V2116 Ophiuchi; [6,2,4,18]). Among the large X-ray binary zoo, we
know so far only another system hosting a neutron star (NS) with a
red giant companion: 4U 1700+24 ([16,9]). GX\,1+4\, shows an
unpredictably variable X-ray flux on timescales from hours to decades.
At the time of its discovery ([13]) it was one of the brightest object
in the X-ray sky and it had the largest spin-up rate recorded for any
pulsar at that time.  The average spin-up trend reversed inexplicably
in 1983 switching to spin-down at approximately the same rate. So far
a remarkable number of changes in the sign of the torque has been
observed for this source ([3]). 

This is somehow a peculiar object among the X-ray binaries not only
because of its red giant companion, but also because of the high
magnetic field that the NS is believed to have
($\sim2-3\times10^{13}$G; [7,11,5]). In fact, the presence of a such
high magnetic field in a slowly rotating NS with a red giant
companion is an intriguing puzzle for the evolutionary scenario of
this system.

Here we report on the timing and spectral X-ray properties of GX 1+4,
in particular on a strange drop in the flux occurred in 2000 November
1st during a {\it BeppoSAX} observation.

\section{Timing and spectral analysis}

The {\it BeppoSAX} observatory covered more than three decades of
energy, from 0.1--200\,keV. The payload was composed by four co-aligned
instruments: the Narrow Field Instruments ([1]: LECS, 0.1--10\,keV;
MECS, 1--10\,keV; HPGSPC, 4--100\,keV; PDS, 15--200\,keV) and the Wide
Field Cameras ([12]). All the four NFI instruments were on during the
$\sim$3.5\,days {\it BeppoSAX} observation carried out around 2000 November 1st.

%%%%%%%%%%%%%%%%%%%%%%%%%%%%%%%%%%%%%%%%%
\begin{figure}[t]
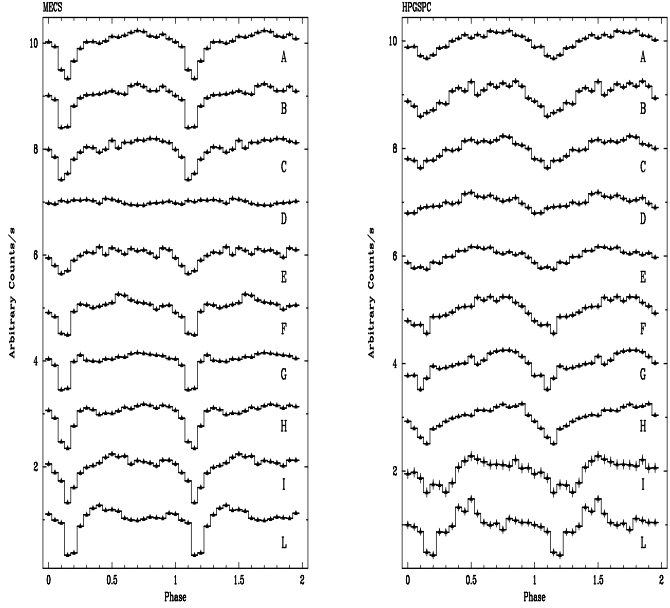

\centerline{
\hbox{
\psfig{figure=me_efold_cospar.ps,width=4.0cm,height=8cm,angle=270} 
\hspace{0.5cm}
\psfig{figure=hp_efold_cospar.ps,width=4.0cm,height=8cm,angle=270}  }}
\caption{MECS (1.65--7\,keV range) and HPGSPC 
(7--35\,keV range) folded lightcurves at the NS spin period
for all time intervals (see also Fig.\,1)}
\end{figure}
%%%%%%%%%%%%%%%%%%%%%%%%%%%%%%%%%%%%%%%%%%%%%%%%%

The lightcurve of the X-ray source, in the 0.1--200\,keV energy range,
showed a large flux variability (see Fig.\,1). We searched for
coherent pulsations performing a power spectrum analysis followed by a
phase-fitting analysis, and we found the spin period value of $P_{s} =
134.925\pm0.001$\,s (phase zero calculated at TJD 11785.000781; errors
in the text are at 1$\sigma$ confidence level). In order to study the
possible evolution or changes of the timing properties of the source,
we divided the observation in 10 time intervals and looked for
pulsations all over each interval in different energy bands. Making
this division we noticed that outside of the low X-ray flux event, all
instruments showed pulsations at the same spin period
$P_{s}=134.925\pm0.001$\,s in the whole {\it BeppoSAX} energy range
(see Fig.\,2), while in interval D no pulsed emission was detected
below $\sim$7\,keV. As we can see from Fig.\,2, the pulse shape was
highly variable either in time or in energy.  Comparing the phase at
which the minimum of the pulse occurs among the HPGSPC profiles, we
found shifts in phase between all curves: e.g. the folded lightcurve
in the interval L is shifted in phase by 0.22+-0.05 with respect to
that in the interval D. Along with the pulse profile changes, the
pulsed fraction also varies. During the event the pulsed fraction
below 7\,keV was consistent with zero (6\% upper limit), while it
increases until $20\pm4$\% in the 7--35\,keV energy range.

In order to fit the spectra of the source in all the intervals, we
first tried several simple models such as a bremsstrahlung emission,
as a blackbody plus a power-law, or as a multicolour blackbody, but
all these models gave a bad chi-square value. Noticing then that the
source spectrum in the low-flux state was exactly what is expected for
a Compton-reflection dominated spectrum, we tried to fit all the
spectra with an absorbed cut-off power law with a reflection component
(pexrav model in {\em Xspec}; [15]).  This was actually the best model
for all the time resolved spectra ($\chi^2_{\nu} \sim 0.98-1.1 $) with
a very small relative reflection strength (refl$\simeq$0--0.3) outside
the event and being $42\pm3$ during the low-flux event (see
Fig.\,3). The $N_{\rm H}$ ranged between
7--48$\times10^{22}$\,cm$^{-2}$, reaching the maximum values just
before and after the event, while at higher energies, the spectrum
showed a hardening when the Compton scattering component started to
dominate, soon after the flux dropped.

%%%%%%%%%%%%%%%%%%%%%%%%%%%%%%%%%%%%%%%%%%%%%%%%%%%%%%%%%%%%%%%%%%%%%%
\begin{figure}[t]
\centerline{\psfig{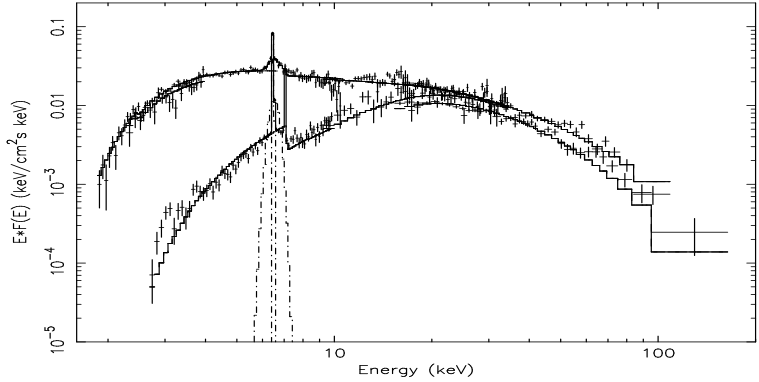} }
\caption{Comparison of the spectrum out and during the low-flux event (the spectra of intervals A and D are the top and bottom data, respectively). Both continuum spectra are fitted with the {\em Xspec} model: wabs*pexrav, but with largely different spectral parameters (see text for details).}
\end{figure}

%%%%%%%%%%%%%%%%%%%%%%%%%%%%%%%%%%%%%%%%%%%%%%%%%

All time-resolved spectra showed at least one emission line. Only one
broad ($\sigma\sim$0.3\,keV) Fe emission line was present at
$E\sim$6.55\,keV outside of the low-flux event, while in the intervals
D and E, the reduced persistent flux revealed the presence of two
narrow lines at $\sim$6.45\,keV and $\sim$7.05\,keV, with very high
equivalent widths ($\sim$2.1 and 0.5\,keV, respectively). The
interpretation of all these lines is very difficult and uncertain
considering the limited energy resolution of the MECS. Thus, we can
only speculate on the origin of the lines. Our idea is that the broad
line at 6.55\,keV might be the blend of the neutral Fe
$K_{\alpha}$ around 6.4\,keV and the ionized Fe XXV at 6.7\,keV,
while during the event only the neutral Fe components are present,
the $K_{\alpha}$ and the $K_{\beta}$.

\section{Discussion} 

We report here on the longest uninterrupted observation of the X-ray
binary system GX\,1+4. The most important results of this analysis
are: i) a Compton-reflection component dominates the source spectrum
during a low-flux emission event, ii) the discovery of the line at
$\sim$7.05\,keV during such low-intensity event, iii) during the
latter, a pulsed X-ray emission was detected only at energies $>$
7\,keV, vi) the detection of a highly variable pulse profile and v)
the detection of a shift in phase of the minimum of the high-energy
pulse profile during the low-flux event.

Hereafter we consider a few models for this event taking in account
the results of our analysis. Our first idea was that the X-ray source
entered in a different emission status (as happen for some X-ray
binaries), where the flux diminishes and the spectrum changes. However,
since one spectral model was able to describe all the spectra and that
the spectrum during the low-flux event was Compton-reflection
dominated, we ruled out this first hypothesis.

In fact, the reflection component should be produced by a Compton thick
material that reprocesses the source photons. This reprocessing occurs
through the Compton-reflection process, where X-rays and $\gamma$-rays
emitted by a source impinge upon a slab of material (e.g. accretion
disk) and re-emerge with a spectrum altered by the Compton scattering
and the bound-free absorption. This process cause a characteristic
hardening in the X-ray spectrum which is due to the onset of the
reflected component, which appears at energies $>$10\,keV, as a result
of the increased importance of the Compton scattering in comparison
with the bound-free absorption, which instead dominates at
low-energies ([21, 14]). At this point, we were then interested to
figure out the geometry and the nature of the material responsible for
the reflection. 

Concerning the geometry, one possibility is that the NS was simply
hidden in a partial eclipse caused by the giant companion. The partial
covering might be due to part of the giant star (a spherical cap)
which temporarily occults the NS direct X-ray emission from our line
of sight.  Although this eclipse scenario is consistent with the
source variability, however, a wide solid angle of Compton thick
material around the source is needed in order to produce such highly
reflection dominated spectrum, and this large solide angle cannot be
produced by the stellar companion wind only. Moreover, taking into
account the large size of the companion star compared with the 10\,km
radius of the NS, the occurrence of a $\sim$90\,ks eclipse, requires
an ad hoc fine tuning of the line of sight inclination with respect to
the orbital plane.

Another possible geometry might be a torus-like accretion disk around
the compact source, due to matter coming from the giant companion.
The lobes of the torus become thicker and increase in volume with
increasing of the accretion rate. The direct emission of the NS would
had been hidden by one side of the torus while the other side of the
torus reflected it, causing the drop of the flux and the Compton
reflected spectrum.

This model can explain as well the lack of detection of the low-energy
pulsations, as they are too weak to keep their coherence through the
scattering in the thick torus material. Moreover, during the
low-emission event the photons make a different path before reaching
the observer, compared to the direct emission phases, and this can be
the cause of the shift in the spin-phases revealed in the high-energy
pulsations (see Fig.\,2 right panel).

This scenario requires an highly variable mass accretion rate,
possibly due to a variable mass loss from the red giant companion,
which is expected from these stars. Thus, note that the accretion
rate variability of this source, is well supported by the detections
of spin-torque variations in the source timing history ([3]).

This latter model we propose in order to explain the low-flux event,
is in analogy with what was proposed for some AGN sources and for some
of the highly absorbed {\it INTEGRAL} sources ([20,19,17]). If this
scenario is correct we can imagine that the highly absorbed {\it
INTEGRAL} sources are compact binary systems of this type for which
our unlucky line of sight, hides the direct emission of the compact
object behind the accretion torus lobes . Note that the absorption
value reached by GX\,1+4 in some part of the observation is similar of
the $N_{\rm H}$ found for some of this highly absorbed {\it INTEGRAL}
sources, although a Compton-reflection component was never
unambiguously revealed in these sources so far.

\vspace{1cm}

NR acknowledges the referee Andrzej Zdziarski for his important
comments and advises, and the Editor Roland Diehl for his patience.


\begin{thebibliography}{}

\bibitem {1}
      Boella, G., et al. 1997, A\&AS, 122, 299
\bibitem {2} 
     Chakrabarty D., et al. 1997a, A\&AS, 190, 450 
\bibitem {3} 
      Chakrabarty D. et al. 1997b, ApJ, 481, L101 
\bibitem {4} 
      Chakrabarty D, et al. 1998, ApJ, 497, L39
\bibitem{5}
      Cui W. et al. 1997,  ApJ, 482, L163 
\bibitem{6}
     Davidsen A., et al. 1977, ApJ,  211 , 866 
\bibitem {7} 
     Dotani T., et al. 1989, PASJ, 41, 427
\bibitem {8} 
     Galloway D.K., et al. 2001, MNRAS, 325, 419
\bibitem {9} 
    Galloway D.K., et al. 2002,  ApJ, 580, 1065
\bibitem {10} 
    Giles A.B., et al. 2000, ApJ, 529, 447
\bibitem {11} 
    Greenhill J.G., et al. 1993, MNRAS, 260, 21
\bibitem {12} 
     Jager R. et al. 1997, A\&AS, 125, 557
\bibitem {13} 
      Lewin W.H.G., Rickter G.R. \& McClintock J.E, 1971, ApJ, 169, L17 
\bibitem {14} 
     Lightman A.P. \& White T.R. 1988, ApJ, 335, 57
\bibitem {15} 
       Magdziarz P. \& Zdziarski A.A. 1995, MNRAS, 273, 837
\bibitem {16} 
       Masetti, N., et al. 2002, A\&A, 382, 104
\bibitem {17} 
      Matt G. \& Guainazzi M., 2003, MNRAS, 341, L13
\bibitem {18} 
      Pereira M.G., et al., 1997, IAUC 6698, 4
\bibitem {19} 
     Revnivtsev M. G., et al. 2003, AstL, 29, 587
\bibitem {20} 
     Walter R.,  et al.. 2003, A\&A, 411, L427
\bibitem {21} 
    White T.R., Lightman A.P. \& Zdziarski A.A., 1988, ApJ, 331, 939

\end{thebibliography}
\end{document}